# Optimal Power Flow Considering Time of Use and Real-Time Pricing Demand Response Programs

Sayyad Nojavan, Vafa Ajoulabadi, Tohid Khalili, *Student Member, IEEE*, Ali Bidram*, Senior Member, IEEE*

***Abstract*—In recent years, the implementation of the demand response (DR) programs in the power system's scheduling and operation is increased. DR is used to improve the consumers' and power providers' economic condition. That said, optimal power flow is a fundamental concept in the power system operation and control. The impact of exploiting DR programs in the power management of the systems is of significant importance. In this paper, the effect of time-based DR programs on the cost of 24-hour operation of a power system is presented. The effect of the time of use and real-time pricing programs with different participation factors are investigated. In addition, the system's operation cost is studied to analyze the DR programs' role in the current power grids. For this aim, the 14-bus IEEE test system is used to properly implement and simulate the proposed approach.**

## NOMENCLATURE

**Indices and Sets**

| | |
|---|---|
| $t, j, i$ | Index of the time |
| $nb, mb$ | System's buses index |
| $N_{bus}, \Omega_b$ | Set of buses |

**Parameters and Constants**

| | |
|---|---|
| $a, b$ | Coefficients for linear load |
| $\rho_0$ | Initial price of electricity |
| $\rho(i)^{max}$ | Maximum price of electricity |
| $\rho(i)^{min}$ | Minimum price of electricity |
| $P_{d0}$ | Initial demand load |
| $P_d$ | Demand load |
| $El(i,i)$ | Self-elasticity |
| $El(i,j)$ | Cross-elasticity |
| $B^{sh}$ | Charging susceptance |
| $B_{nbmb}$ | Branch nbmb susceptance |
| $G_{nbmb}$ | Branch nbmb conductance |
| $a_C, b_C, c_C$ | Coefficient of operation cost function |
| $S_{max}$ | Power flow limitations |
| $Q_{Cmax}$ | Power generation unit maximum limitation |
| $Q_{Cmin}$ | Power generation unit minimum limitation |
| $\lvert V\rvert_{max}$ | Maximum voltage magnitudes |
| $\lvert V\rvert_{min}$ | Minimum voltage magnitudes |

**Functions and Variables**

| | |
|---|---|
| $\rho$ | Electricity price |
| $P_C$ | Power generation |
| $Q_C$ | Reactive power generation |
| $P_{nbmb}$ | Active power flow of branch nbmb |
| $Q_{nbmb}$ | Reactive power flow of branch nbmb |
| $P_{dnb}$ | Bus nb active demand |
| $Q_{dnb}$ | Bus nb reactive demand |
| $\lvert V_{nb}\rvert$ | Bus nb voltage magnitude |
| $\theta_{nbmb}$ | Branch nbmb admittance angle |
| $Opcost$ | System operation cost function |

Tohid Khalili and Ali Bidram are supported by the National Science Foundation EPSCoR Program under Award #OIA-1757207.
Sayyad Nojavan and Vafa Ajoulabadi are with the Department of Electrical Engineering, University of Bonab, Bonab, Iran. e-mails: (sayyad.nojavan@ubonab.ac.ir, vafa225@gmail.com). Tohid Khalili and Ali Bidram are with the Department of Electrical and Computer Engineering, University of New Mexico, Albuquerque, USA. (e-mails: {khalili, bidram}@unm.edu).

## I. INTRODUCTION

ENVIRONMENTAL issues are one of the significant problems of the current era. Considering the governments obligation to use renewable resources, numerous uncertain parameters are added to the operation of the power system. Thus, operation problems become more challenging. With the restructuring of power systems and the advancement of telecommunication technology, the use of demand response (DR) programs is considered a trustable way to manage the power system appropriately [1]. One of the main concepts of the operation in the power system is to have the optimal power flow (OPF). By solving the OPF problem, the status of the whole system is determined. Some types of OPF are as follows: dynamic OPF, static OPF, security-constrained OPF, deterministic OPF, transient stability-constrained OPF, stochastic OPF, AC OPF, probabilistic OPF, DC OPF, and mixed AC/DC OPF [2]. Regarding the current situation of the power system, it is impossible to ignore the DR programs effect in the scheduling and operation of systems.

Demand-side management (DSM) was initially proposed with the aim of reducing energy consumption, but after the privatization and restructuration of the power systems, several other objectives are considered in the DSM projects. After the restructuring of the power systems and the creation of the electricity market, the DSM was re-created in the form of bilateral contracts, which is known as DR.

The ability of household, commercial, and industrial subscribers to improve their electricity consumption patterns to improve the grid reliability and to achieve reasonable prices is called DR [3]. DR program's goals are divided into short and long-term ones. The long-term goals of DR programs are to defer the need to develop generation capacity and install new power lines, and the short-term goals are to increase network reliability and prevent price spikes. The major advantages of using the DR programs are the competitiveness of the electricity market, risk reduction, proper interaction between supply and demand, the connection between retail and wholesale energy markets, creation of a new tool for customer load management, and the existence of environmental benefits due to reduced usage of the fossil resources.

Both consumers and operators should spend money and invest to take advantage of the DR programs. Consumers' pays could include the installation of new technologies to control energy consumption, installation of distributed generation sources, fuel costs, repair, maintenance, and network connection equipment to distributed generation sources. The costs of the implementers of these programs include the installation of advanced two-way metering equipment for measuring, exchanging, and storing information. Also, when the incentive-based DR program is implemented the system operator should pay incentives to customers.

DR programs can be divided into two main categories: incentive-based and time-based programs. In the incentive-based programs, customers change the amount of their demanded load by considering the contracts, rewards, and penalties. Moreover, the incentive-based programs are divided into six categories: interruptible/curtailable (I/C) service, direct load control (DLC), demand bidding/buy back (DB), ancillary service (A/S) markets, emergency dr program (EDRP), and capacity market program (CAP). On the other side, the time-based programs have no penalty or incentive for their customers, and energy prices are charged to customers at different time intervals. Time- based programs can be divided into three groups: critical peak pricing (CPP), time of use programs (TOU), and real-time pricing (RTP). Fig. 1 shows the general scheme of DR programs [4].

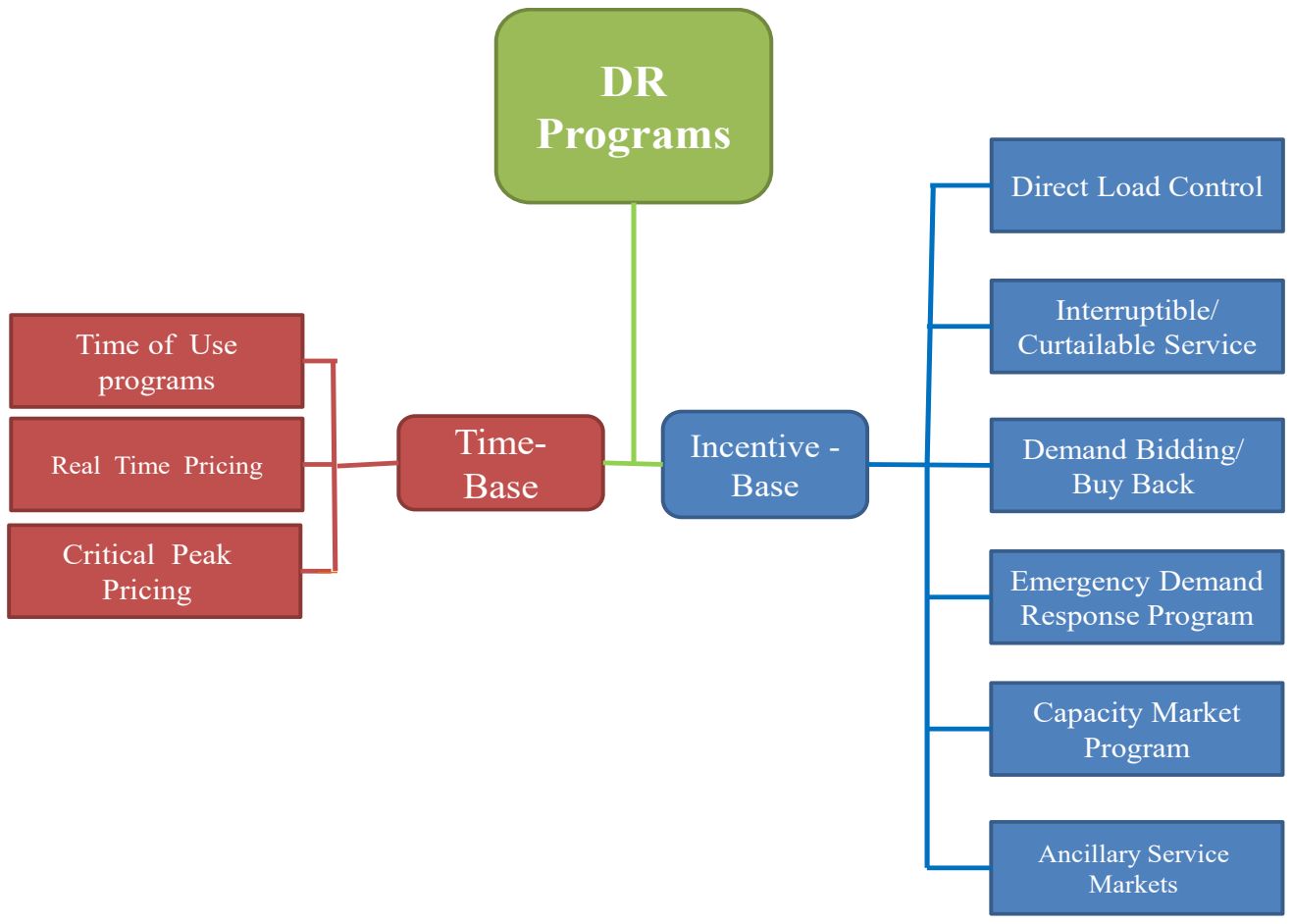

Fig. 1. DR classification.

### A. Literature Review

Many studies and researches have been done in the field of power system's scheduling and operation. But, several challenges and issues have remained that need to be addressed. For instance, the management of a distribution system with distributed generation resources is investigated in [5]. The objective of the [5] is to minimize the previous day's operating cost which is done by optimally controlling the active elements of the network, distributed generation sources, and responsive loads. Also, both incentive and time-based programs have been mathematically modeled [6]. Reference [7] introduces a priority indicator for consumers that participate in the electricity market. This index is determined based on the number and amount of applications to participate in the DR program. Low priority customers with smart energy management permission can use electricity when energy prices are cheap. The results show that the intelligent energy management system is successful in decreasing peak load. Additionally, the load reduction during peak times depends on the amount of participation in the DR program during peak hours.

In [8], a time-based DR program that includes TOU, actual pricing, and peak time pricing is modeled and implemented. Also, these programs have been compared with different participation factors. In [4], the focus is on two types of incentive-based DR programs. In this reference, several scenarios with different rewards and penalties are introduced and the simulation results of these scenarios are analyzed and compared [4]. In [9], DR programs have been used to reduce the operating cost of the energy hub. In [10], optimal microgrid's scheduling has been studied with the implementation of the incentive-based DR program considering the uncertainty of renewable energy sources and load. The performance and efficiency of renewable energy sources and storages have been studied from the uncertainty point of view in [11]. In [12] time-based DR program for commercial, residential, academia, and industrial load curves is implemented.

In this paper, the effect of RTP and TOU DR Programs on the power system 24 hours operating cost is investigated. The effect of various participation factors on the OPF problem has also been studied. And, the obtained results for all scenarios are presented and discussed.

### B. Paper Organization

In this work, section II presents the load modeling with the time-based DR program. In section III, 24-hour OPF problem formulation is explained. Information needed to solve the problem and simulation results are presented in section IV. Section V expresses the conclusion of the proposed approach.

## II. Load Modeling

### A. Elasticity

Load's elasticity is the load sensitivity with respect to the price of the power. The mathematical model of the elasticity is shown by (1). The load elasticity parameter is divided into two parts: self-elasticity and cross-elasticity. Self-elasticity is always negative and cross-elasticity is positive [4]. Equations (2) and (3) represents the mathematical model of the self-elasticity and cross-elasticity, respectively. In this paper, the linear load model is considered and (4) shows the utilized mentioned linear model. Equation (5) is obtained by considering the load's elasticity definition and linear load model [4]. The relationship between load, price, and elasticity is shown by (6). For the daily operation, the desired relation can be represented as a 24-by-24 matrix, which is presented in (7). In this matrix, the elements of the principal diameter are the self-elasticity, and the other elements of the matrix are the cross elasticity [13].

$$El = \frac{\rho_0}{P_{d0}} \frac{\partial P_d}{\partial \rho} \tag{1}$$

$$El(i,i) = \frac{\rho_0(i)}{P_{d0}(i)} \frac{\partial P_d(i)}{\partial \rho(i)} \tag{2}$$

$$El(i,j) = \frac{\rho_0(j)}{P_{d0}(i)} \frac{\partial P_d(i)}{\partial \rho(j)} \quad i \neq j \tag{3}$$

$$P_d = a + b.\rho \tag{4}$$

$$El = b.\frac{\rho_0}{P_{d0}} \tag{5}$$

$$\Delta P_d = El.\Delta\rho \tag{6}$$

$$\begin{bmatrix} \Delta Pd(i) \\ \vdots \\ \Delta Pd(j) \end{bmatrix} = \begin{bmatrix} El(1,1) & \dots & El(1,24) \\ \vdots & El(i,j) & \vdots \\ El(24,1) & \dots & El(24,24) \end{bmatrix} \times \begin{bmatrix} \Delta\rho(i) \\ \vdots \\ \Delta\rho(j) \end{bmatrix} \tag{7}$$

### B. Responsive Loads Modeling

Responsive loads can be divided into two categories: multi-period and single-period loads. The single period load model is not able to be transmitted at other time intervals. This type of load can only be turned on or off. In this type of load, it is impossible to change the time period of the demanded load. Thus, in the i[th] period of the operation, the amount of changes in the consumer's load is obtained from (8) and the profit is obtained from (9). Equation (10) is used to maximize the customer's profitability. After simplification of (10); the final formula is shown in (11). However, multi-period loads can be transferred at different time intervals and can be used at different times of the day. The model of this type of load is calculated based on cross elasticity by (12). The final load model for a responsive load is obtained from (13) which is achieved by combining two single-period and multi-period models [13].

$$\Delta P_d(i) = P_d(i) - P_{d0}(i) \tag{8}$$

$$S\left(P_d(i)\right) = B\left(P_d(i)\right) - P_d(i).\rho(i) \tag{9}$$

$$\frac{\partial S\left(P_d(i)\right)}{\partial P_d(i)} = 0 \tag{10}$$

$$P_d(i) = P_{d0}(i)\left\{1 + \frac{El(i,i)\left[\rho(i) - \rho_0(i)\right]}{\rho_0(i)}\right\} \tag{11}$$

$$P_d(i) = P_{d0}(i) + \sum_{\substack{j=1 \\ j\neq i}}^{24} El(i,j)\frac{P_{d0}(i)}{\rho_0(j)}\left[\rho(j) - \rho_0(j)\right] \tag{12}$$

$$P_d(i) = P_{d0}(i)\left\{1 + \frac{El(i,i)\left[\rho(i) - \rho_0(i)\right]}{\rho_0(i)} + \sum_{\substack{j=1 \\ j\neq i}}^{24} El(i,j)\frac{\left[\rho(j) - \rho_0(j)\right]}{\rho_0(j)}\right\} \tag{13}$$

## III. Problem Formulation

### A. Objective Function

As mentioned, it is impossible to operate power networks without solving the OPF problem. The OPF problem is an optimization problem for minimizing the power grid's operating costs. Equation (14) shows the operating cost of the conventional generation resources, which is modeled as a quadratic function with respect to fuel consumption.

$$Opcost_t = a_C P_{C,t}^{\ 2} + b_C P_{C,t} + C_{C,t} \tag{14}$$

### B. Constraints

The constraints of the optimization problem are related to the network and power generation resources limits. Equations (15) and (16) show the typical AC power flow equations. Equations (17) and (18) present the balance constraints between consumption and generation for active power and reactive power, respectively. By combining (17) and (13), the new active power balance constraint is shown in (19). The line constraints for power transmission are presented in (20). Constraints for bus voltage are shown in (21)-(23). Power generation resources also have their own limitations; Equations (24) and (25) indicate the electricity generation resources' active and reactive power constraints, respectively.

$$P_{C,t,nbmb} = \left|V_{t,bn}\right|^2 G_{nbmb} - \left|V_{t,bn}\right|\left|V_{t,bm}\right|\begin{pmatrix} G_{bnbm}\cos\theta_{t,nbmb} \\ +B_{nbmb}\sin\theta_{t,nbmb}\end{pmatrix} \tag{15}$$

$$\begin{aligned} Q_{C,t,nbmb} = &-\left|V_{t,bn}\right|^2\left(B_{nbmb} + B_{nbmb}^{sh}\right) \\ &-\left|V_{t,bn}\right|\left|V_{t,bm}\right|\left(G_{nbmb}\sin\theta_{t,nbmb} - B_{bnbm}\cos\theta_{t,nbmb}\right) \end{aligned} \tag{16}$$

$$P_{C,t} - P_{t,load} = \sum_{mb\in\Omega_b} P_{t,nbmb} \tag{17}$$

$$Q_{C,t} - Q_{Ct,load} = \sum_{mb\in\Omega_b} Q_{t,nbmb} \tag{18}$$

$$\sum_{mb\in\Omega_b} P_{t,nbmb} = P_{C,t} - \left[P_{d0}(i)\left\{\begin{aligned} &1 + \frac{El(i,i)\left[\rho(i) - \rho_0(i)\right]}{\rho_0(i)} \\ &+\sum_{\substack{j=1 \\ j\neq i}}^{24} El(i,j)\frac{\left[\rho(j) - \rho_0(j)\right]}{\rho_0(j)}\end{aligned}\right\}\right] \tag{19}$$

$$\sqrt{\left[\left(P_{t,nbmb}\right)^2 + \left(Q_{t,nbmb}\right)^2\right]} \le S_{max,nbmb} \tag{20}$$

$$\left|V\right|_{min} \le \left|V\right|_{nb,t} \le \left|V\right|_{max} \tag{21}$$

$$\left|V\right|_{nb,t} = 1 \qquad nb \in N_{bus} \tag{22}$$

$$\theta_{min} \le \theta_{bn,t} \le \theta_{max} \text{ or } \theta_{t,n} = 0 \qquad bn \in N_{bus} \tag{23}$$

$$P_{C,min} \le P_{C,t} \le P_{C,max} \tag{24}$$

$$Q_{C,min} \le Q_{C,t} \le Q_{C,\,max} \tag{25}$$

## IV. Test System And Simulation Results

The IEEE standard 14-bus network, shown in Fig. 2, is used to perform the simulations [14]. For all cases in the considered system, the 24-hour load profile is used which is indicated in Fig. 3 [8]. The required coefficients of the system's operating cost for (14) are demonstrated in Table I [14].

Not all customers are willing to participate in DR programs. The value of the participation factor should be used as a parameter to show the amount of the customers' participation factor and related loads. To study the effect of this parameter, the value of the participation coefficient is considered as 10%, 20%, and 50%. Also, the assumed self-elasticity and cross elasticity are shown in Table II [13]. The TOU program prices are 30, 70, and 120 \$/MWh for the valley period, off-peak, and peak hours, respectively. In addition, 30, 50, 70, 100, and 120 \$/MWh are considered in the RTP program.

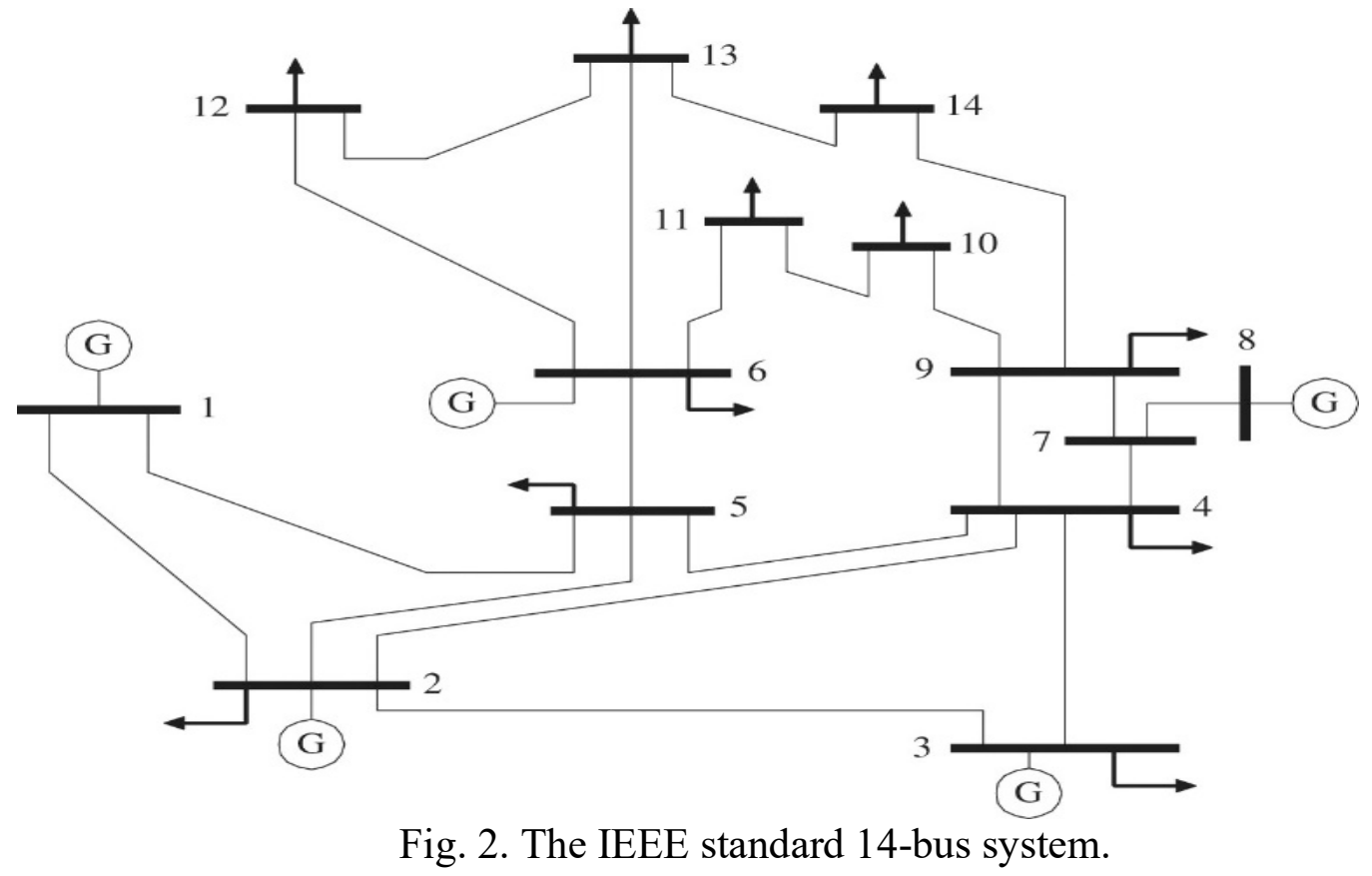

Fig. 2. The IEEE standard 14-bus system.

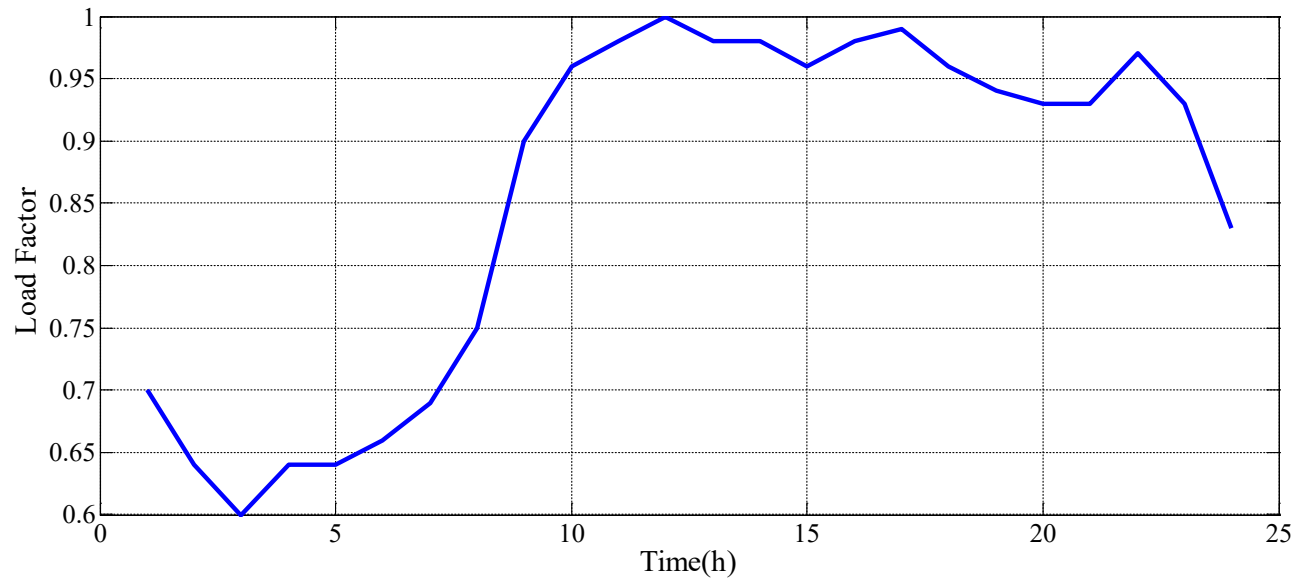

Fig. 3. 24-hour load profile.

TABLE I
COEFFICIENTS OF OPERATING COST

| Bus | a | b | c |
|---|---|---|---|
| 1 | 0.043 | 20 | 0 |
| 2 | 0.25 | 20 | 0 |
| 3 | 0.01 | 40 | 0 |
| 6 | 0.01 | 40 | 0 |
| 8 | 0.01 | 40 | 0 |

TABLE II
CROSS-ELASTICITY AND SELF-ELASTICITY

| # | Peak period | Off-peak period | Valley period |
|---|---|---|---|
| Peak period | -0.1 | 0.016 | 0.012 |
| Off-peak period | 0.016 | -0.1 | 0.01 |
| Valley period | 0.012 | 0.01 | -0.1 |

Implementation of the DR program is performed in the MATLAB software. The effect of RTP and TOU programs on the considered load curve can be seen in Figs. 4 and 5, respectively. In each figure, the participation factors are 10%, 20%, and 50%. As Figs. 4 and 5 show, by increasing the participation factor the peak of the load profile is disappear and shaved. Regarding the fact that which DR programs are utilized, different profiles are obtained. Indeed, the output and impact of these two considered programs vary by changing the participation factor. Also, the OPF problem's objective is the minimization of the operation cost. The network operator must perform AC OPF calculations, which is a nonlinear optimization problem. The MATPOWER toolbox in MATLAB has been used to simulate and solve the proposed optimization problem [15].

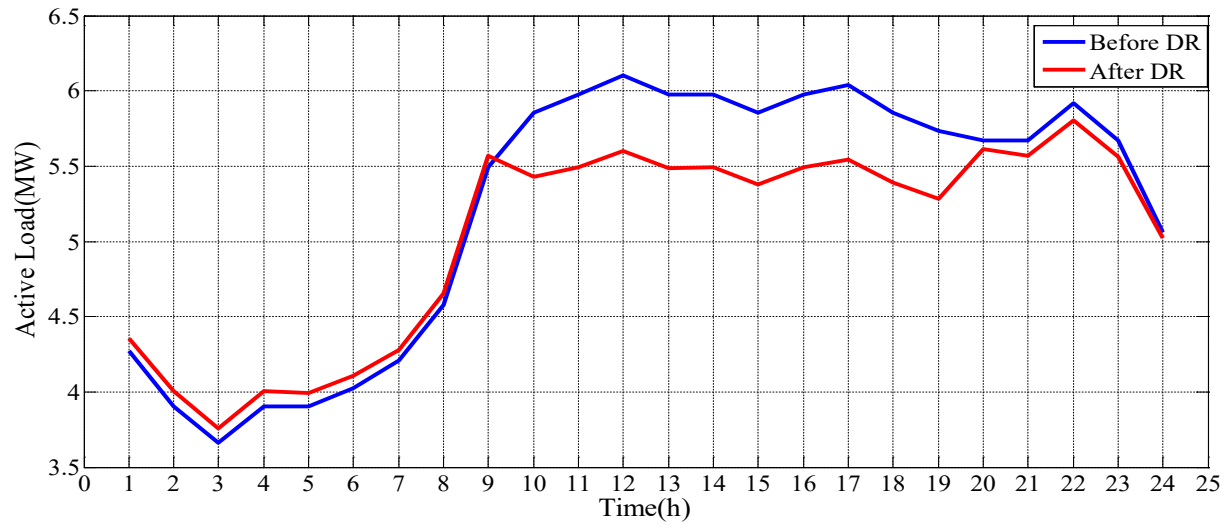

(a)

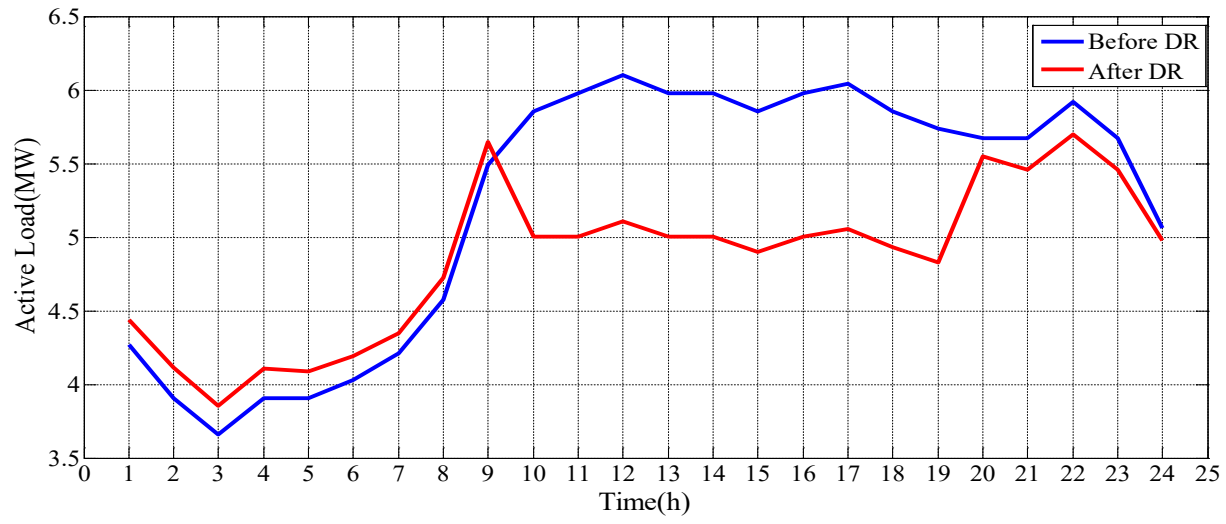

(b)

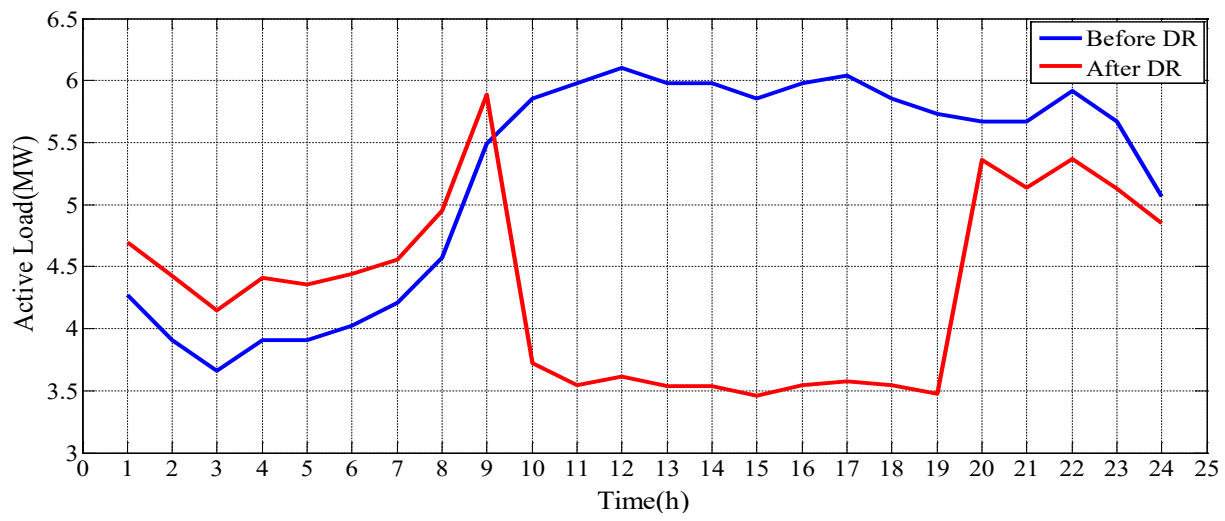

(c)

Fig. 4. The impact of the RTP program on the considered load profile with participation factor of (a) 10%; (b) 20%; (c) 50%.

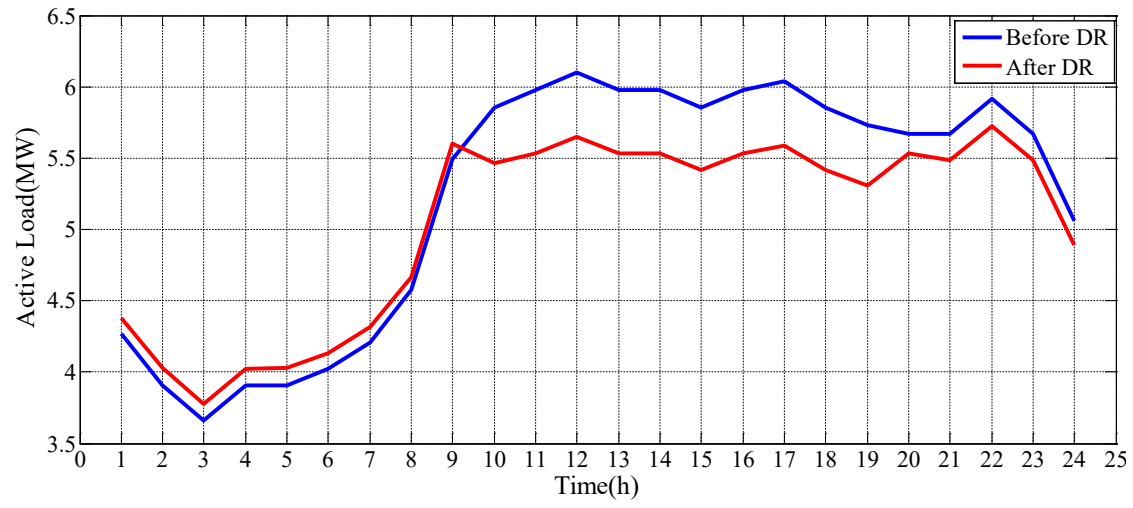

(a)

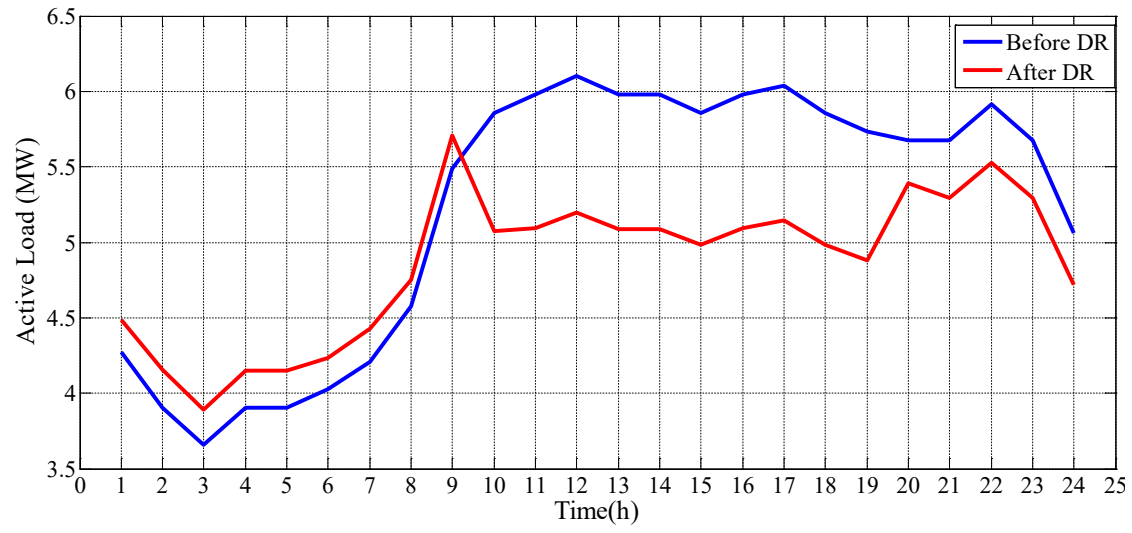

(b)

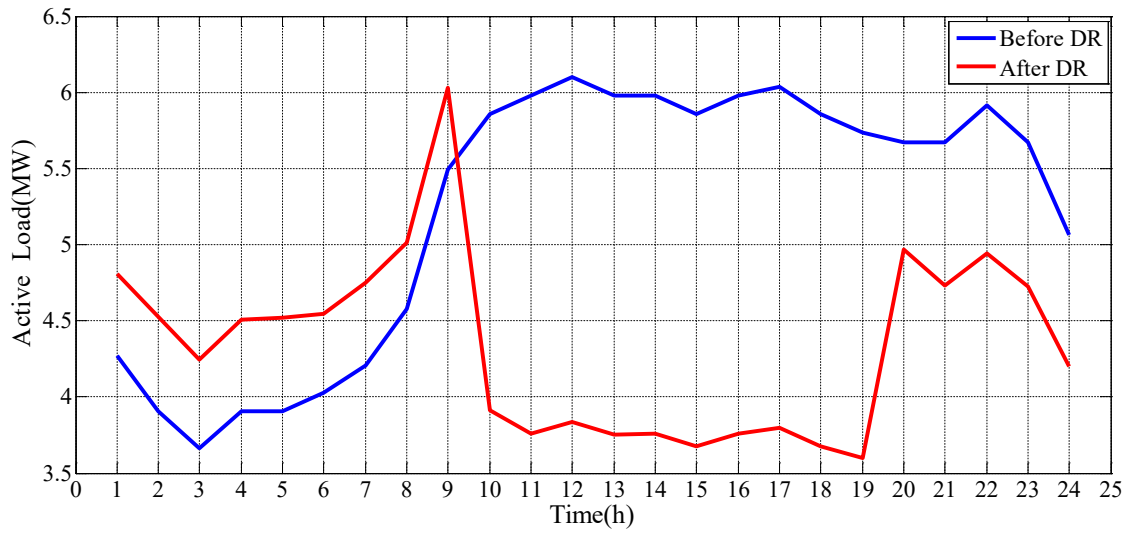

(c)

Fig. 5. The effect of the TOU program on the considered load profile with participation factor of (a) 10%; (b) 20%; (c) 50%.

Fig. 6 shows the system's daily operation cost before the implementation of the DR programs. Moreover, Figs. 7 and 8 illustrate the network operation cost in the 24 hours after the implementation of the RTP and TOU programs. To show the effect of participation factors on the operating cost, DR programs with different participation factors of 10% (Fig. 7a and Fig. 8a), 20% (Fig. 7b and Fig. 8b), and 50% (Fig. 7 c and Fig. 8c) are considered. By considering the Figs. 7 and 8, it is proved that by DR program implementation the operating cost of the system is dramatically reduced. In addition, when the operator and consumer increase the participation factor, the operating cost is significantly decreased. Moreover, each of the used DR programs has a different impact on the operating cost of the system which is dependent on the participation factor of the consumers.

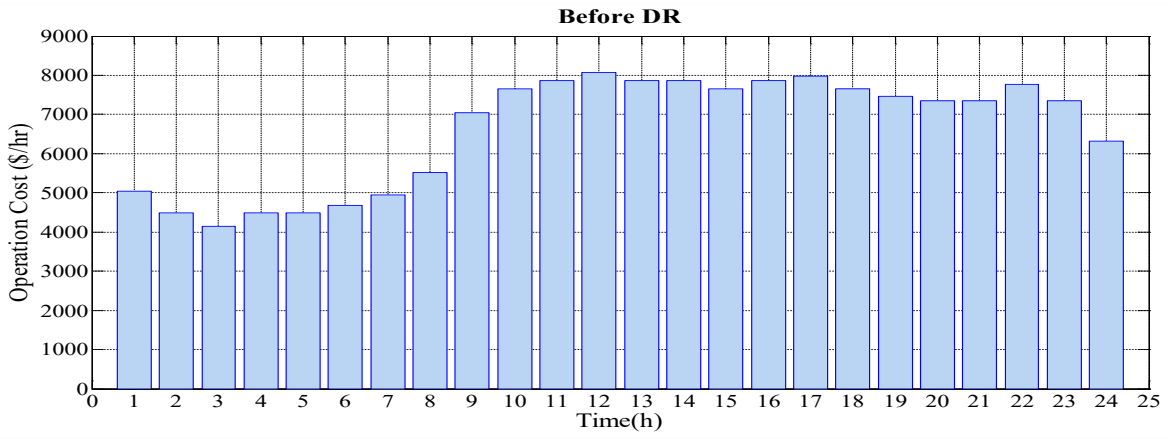

Fig. 6. The network operation cost in the 24 hours before the implementation of the RTP and TOU programs.

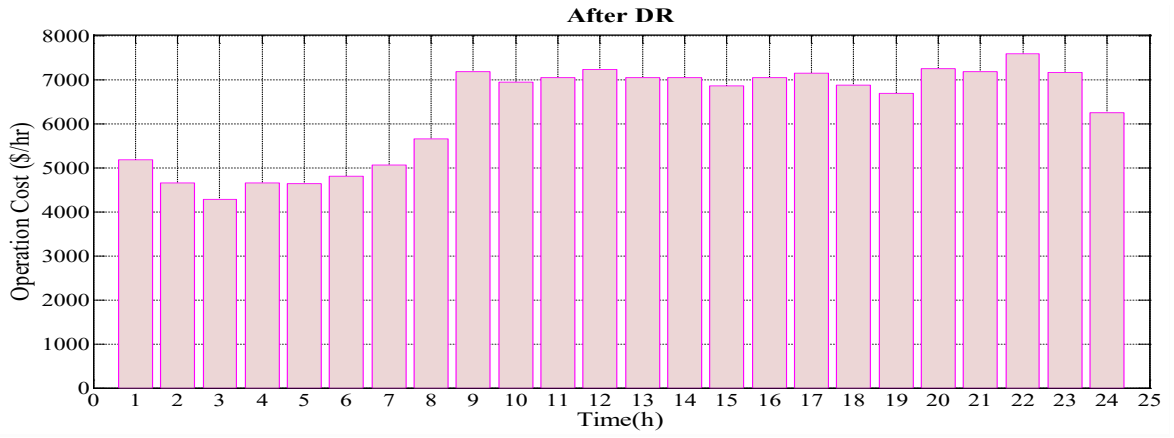

(a)

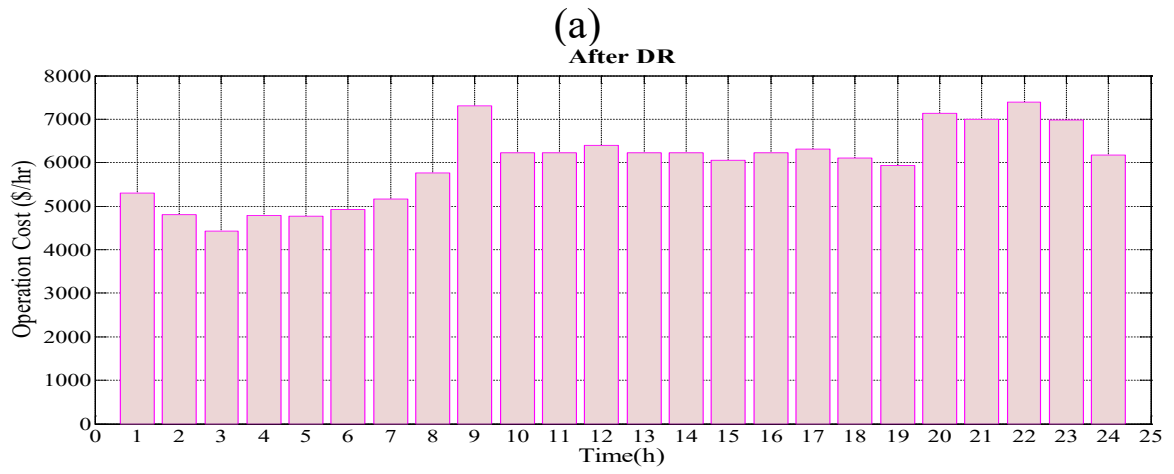

(b)

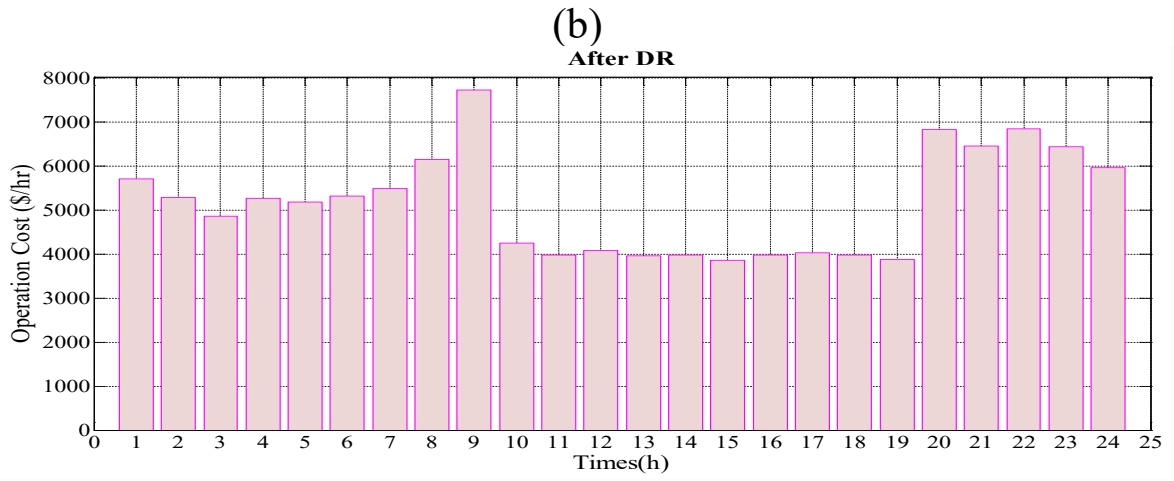

(c)

Fig. 7. The network operation cost in the 24 hours after the implementation of the RTP program with participation factor of (a) 10%; (b) 20%; (c) 50%.

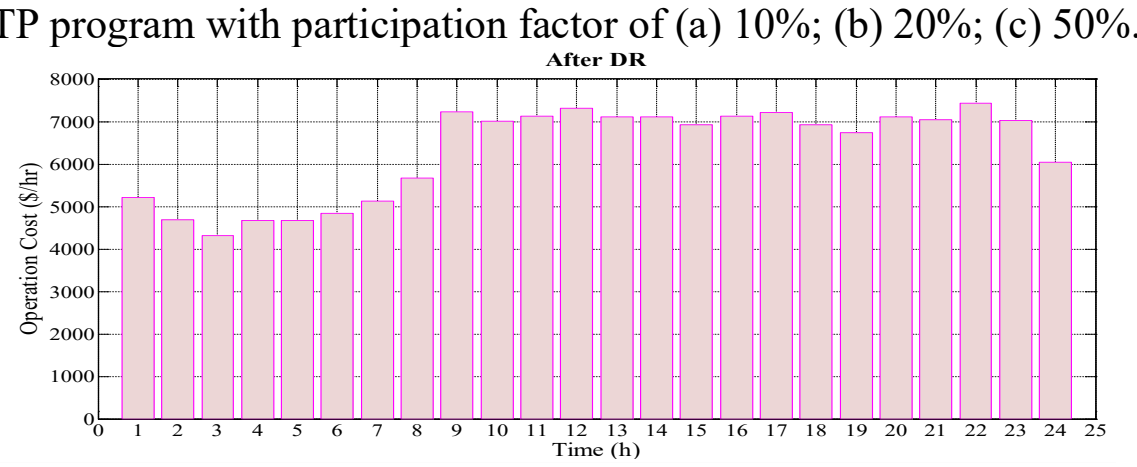

(a)

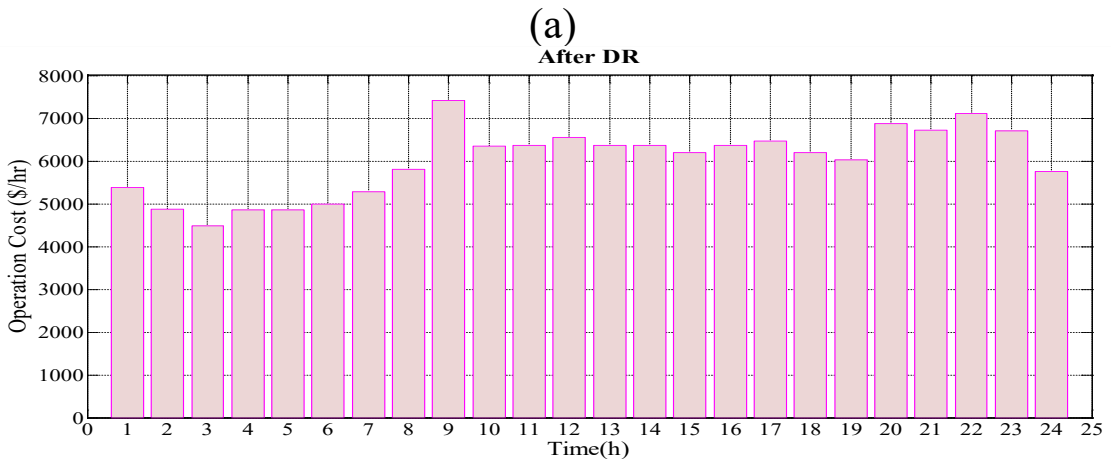

(b)

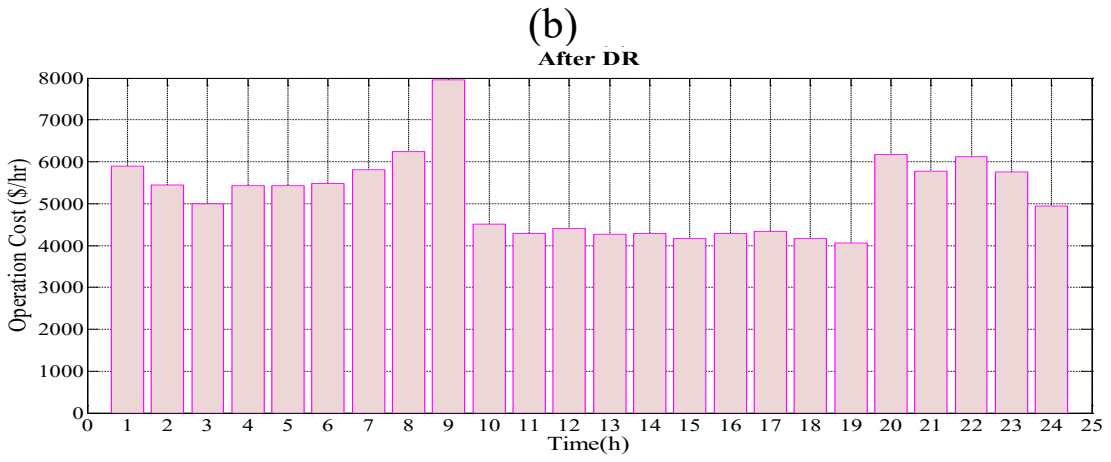

(c)

Fig. 8. The network operation cost in the 24 hours after the implementation of the TOU program with participation factor of (a) 10%; (b) 20%; (c) 50%.

## V. Conclusion

In this study, RTP and TOU DR programs are implemented on the IEEE standard 14-bus system. The effect of these DR programs on the load profile is investigated. In addition, the operation cost of the system is calculated considering different cases. According to the simulation results, DR programs make the 24-hour load profile smoother. In the peak periods, the network operator will have to turn on its pricy units. Thus, reducing the consumption in peak hours and transferring the demand to the valley and off-peak intervals dramatically decrease the operating costs of the network in 24 hours. Furthermore, various participation factors will also have a large impact on reducing operating costs and smoothing the load profile, especially during peak hours. In future work, the reliability and robustness of the power system could be analyzed under the same considered cases.